\documentclass[epjCONF]{svjour}
\usepackage{graphics}
\usepackage[varg]{txfonts} 
\usepackage[latin1]{inputenc}
\session-title{Hadron Collider Physics Symposium 2011}
\begin{document}
\title{Measurement of the production cross section for $Z/\gamma^*$ in
association with jets in pp collisions at $\sqrt{s}$ = 7 TeV with the ATLAS
Detector}

\author{Evelin Meoni \inst{1}\fnmsep\thanks{\email{evelin.meoni@cern.ch}} on behalf of the ATLAS Collaboration}
\institute{\inst{1}Institut de F\'{i}sica d'Altes Energies (IFAE). Edifici Cn, Universitat Aut\`{o}noma de Barcelona (UAB), E-08193 Bellaterra (Barcelona), Spain.}
\abstract{
We present results on the production of jets of particles in association with a $Z/\gamma^*$ boson, in proton-proton collisions at $\sqrt{s} = 7$~TeV with the ATLAS detector at the LHC. The analysis includes the full 2010 data set, collected with a low rate of multiple proton-proton collisions in the accelerator, corresponding to an integrated luminosity of  $36 \ \rm pb^{-1}$. Inclusive jet cross sections in $Z/\gamma^*$ events, with $Z/\gamma^*$ decaying into electron or muon  pairs, are measured for jets with transverse momentum $p_T >$~30~GeV and jet rapidity $|y| < 4.4$.  The measurements are compared to next-to-leading-order perturbative QCD calculations, and to predictions from different Monte Carlo generators implementing leading-order matrix elements supplemented by parton showers.
} 
\maketitle
\section{Introduction}
\label{intro}
The measurement of the cross sections for the production of hadronic jets in association with a $Z/\gamma^*$ boson
at LHC  provides
a stringent test of perturbative quantum chromodynamics (pQCD). Moreover, since these processes form important
backgrounds for searches of new physics,
 their detailed measurement is a first step in 
the  discovery program at LHC. 
\\
This contribution presents a review of the   measurements of jet production  in events with a  $Z/\gamma^*$ boson
 using $36 \ \rm pb^{-1}$ 
of data collected by the ATLAS experiment~\cite{ref:atlas} in 2010 
at $\sqrt{s}$ = 7 TeV.
During this period, the luminosity of the machine has grown roughly exponentially with running
time up to 2.1 $\times$ 10$^{32}$ cm$^{-2}$ s$^{-1}$. The ATLAS detector
performed well throughout the 2010 run and its response was quickly understood.\\
The measurements, described in detail in~\cite{ref:PAPER}, are compared with the available next-to-leading-order (NLO) pQCD
calculations, providing a validation of the theory in the new kinematic regime,
and with the Monte Carlo (MC) predictions that include leading-order (LO) matrix elements
supplemented by parton showers. The latter are affected by large scale uncertainties and need to be tuned and
validated using data.  
\section{Event Selection and Background estimation}
\label{sec:1}
Events are required to have a reconstructed primary vertex  with at least
3 tracks associated to it.
Events are selected with a $Z/\gamma^*$ boson decaying into a pair of electrons ($e^{+}e^{-}$) 
or muons ($\mu^{+}\mu^{-}$). In both cases a single lepton trigger, electron or muon, is employed.
In the electron channel, the events are  selected  to have two oppositely charged reconstructed electrons 
 with transverse energy E$_T >$  20 GeV, pseudorapidity in the range $|\eta_{e}| <$ 2.47 (where the
transition region between calorimeter sections 1.37 $< |\eta_{e}| <$ 1.52 is excluded).
In the muon channel, the events are  selected to have two oppositely charged reconstructed
muons with transverse momentum p$_T >$  20 GeV, pseudorapidity in the range $|\eta_{\mu}| <$ 2.4.
In both cases, a dilepton invariant mass in the
range 66 GeV$< m_{ll} <$ 116 GeV
is required. 
Jets  are reconstructed using the anti-kt algorithm~\cite{ref:jet_rec} with a distance
parameter R=0.4. The inputs to the anti-kt jet algorithm are clusters of calorimeter
cells seeded by cells with energy that is significantly above the measured noise.
The measured jet transverse momentum p$_T$ is corrected to the particle level scale using an average correction,
computed as a function of jet  transverse momentum and pseudorapidity, and extracted
from MC simulation. 
In this analysis, jets are selected with corrected p$_T >$ 30 GeV and $|y|<$ 4.4 
separated from each of the two leptons by a $\Delta R >$0.5.\\
The background contamination is estimated using MC simulated samples, except for
the QCD multijet contribution, where a data-driven method is employed for each channel. A template fit is used in the electron channel,
whereas the di-muon mass versus the muon isolation plane  is employed in the muon case.
 In the electron channel, the total background increases from 5$\%$ to 17$\%$ as the inclusive jet multiplicity (N$_{jet}$) increases and is
dominated by multi-jet processes, followed by $t\bar{t}$ and diboson processes. In the muon channel, the background  increases from 2$\%$ to 10$\%$ as $N_{jet}$ increases,
dominated by $t\bar{t}$ and diboson processes.
\section{Cross Section Measurement}
\label{sec:2}
The jet measurements are corrected for detector effects back to the particle level using a bin-by-bin correction
procedure, based on ALPGEN~\cite{ref:alp} MC simulated samples, that corrects for jet selection efficiency and resolution effects and also
accounts for the efficiency of the $Z/\gamma^*$ identification. At particle level, the lepton kinematics include the
contributions from the photons radiated within a cone of radius 0.1 around the lepton direction.\\
The total cross sections are measured as functions of N$_{jet}$  in the fiducial region
and the
inclusive jet differential cross sections are measured as functions of jet
 p$_T$  and  $|y|$.
The differential cross sections are also measured as functions of p$_T$ and
 $|y|$ of the leading jet (highest p$_T$) and second leading jet in Z/$\gamma^*$ events with at least one and two jets in the final
state, respectively. For the latter, the cross section is measured also as a function of the invariant mass and the angular
separation of the two leading jets.\\
The measured differential cross sections are defined as function of a given
$\xi$ as:
\begin{equation}
\frac{d\sigma}{d\xi }= \frac{1}{L} \frac{1}{\Delta \xi} (N_{Data} - N_{backg}) \times U(\xi)
\end{equation}
where, for each bin in $\xi$, $N_{Data}$ and $N_{backg}$ denote the number of entries (events or jets) observed in data and the
background prediction, respectively, and $\Delta \xi$ is the bin width, $U(\xi)$ is the correction factor, and $L$ is the total integrated luminosity.\\

A detailed estimation of the systematic uncertainties that affect the measurement has been carried out. In particular the impact of uncertainties related to the
jet energy scale, the jet energy resolution, the background estimation,  the leptons and the unfolding have been evaluated. 
An additional 3.4$\%$ uncertainty on the total integrated luminosity is
also taken into account.
The main systematic source is the jet energy scale that increases from
7$\%$ to 22$\%$ as N$_{jet}$ increases and from 8$\%$ to 12$\%$ as the p$_T$ increases.
The total systematic
uncertainty increases
from 9$\%$ to 23$\%$ as
N$_{jet}$ increases; and
from 10$\%$ at low p$_T$
to 13$\%$ at high p$_T$.
\section{Theoretical Predictions}
\label{sec:3}
The cross section results are compared to the NLO pQCD predictions, as computed with
BlackHat~\cite{ref:BH} using CTEQ6.6 PDFs~\cite{ref:PDF} and factorization and renormalization scale $\mu = H_T/2$ ($H_T$ is the scalar sum of the
p$_T$ of all particles) where the predictions include non-perturbative corrections.\\
The measurements are also compared to the LO predictions including parton shower, as determined in
ALPGEN~\cite{ref:alp}, Sherpa~\cite{ref:Sh} and PYTHIA~\cite{ref:Ph}.
The ALPGEN and Sherpa samples are normalized to the next-to-next-to-leading order (NNLO) pQCD inclusive Drell-Yan prediction. The PYTHIA sample has been normalized to data
from the average of electron and muon cross section in the $\ge 1$  jet bin.
\section{Results}
\label{sec:4}
\begin{figure}
\begin{center}
\resizebox{0.75\columnwidth}{!}{%
\includegraphics{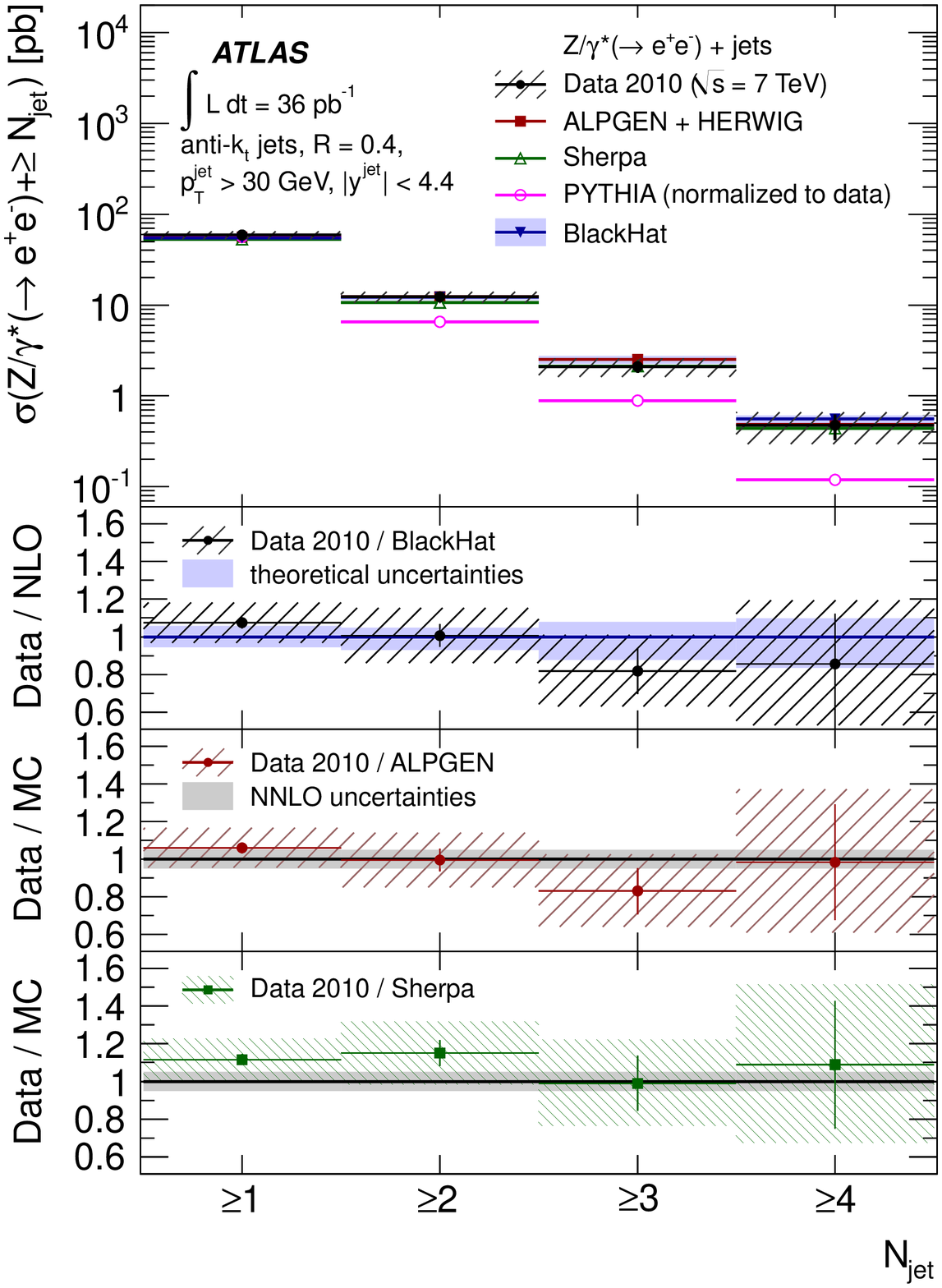} }
\end{center}
\caption{Measured cross section  for Z/$\gamma^*(\to e^+e^-)$+jets production
as a function of the inclusive jet multiplicity. In this and in figures \ref{fig:2}-\ref{fig:4}  the error bars indicate the statistical uncertainty and the dashed areas the statistical and
systematic uncertainties added in quadrature. The measurements are compared to NLO pQCD predictions from BlackHat and to  MC predictions from ALPGEN, Sherpa and PYTHIA.}
\label{fig:1}
\end{figure}

\begin{figure}
\begin{center}
\resizebox{0.75\columnwidth}{!}{%
\includegraphics{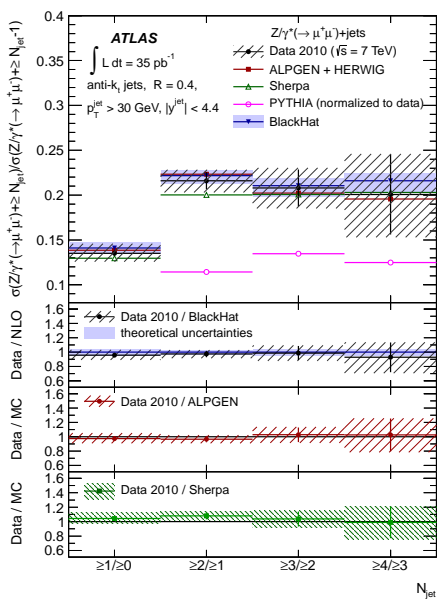} }
\end{center}
\caption{Ratio of cross sections for Z/$\gamma^*(\to \mu^+\mu^-)$+jets production as a function of the inclusive jet multiplicity.}
\label{fig:2}
\end{figure}

\begin{figure}
\begin{center}
\resizebox{0.75\columnwidth}{!}{%
\includegraphics{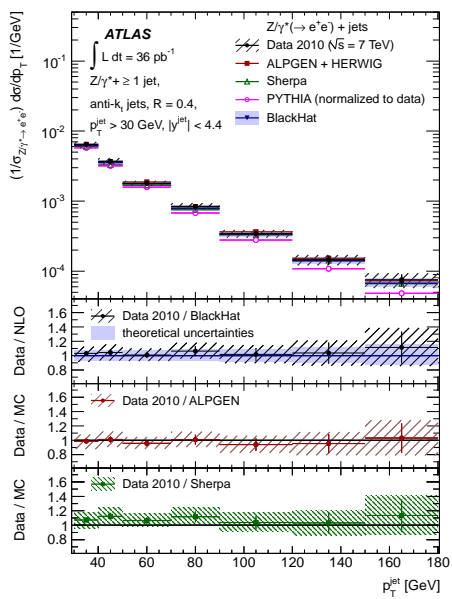} }
\end{center}
\caption{Inclusive jet cross section in  Z/$\gamma^*(\to e^+e^-)$+jets production
normalized by Drell-Yan cross section as a function of p$_T$.}
\label{fig:3}
\end{figure}

\begin{figure}
\begin{center}
\resizebox{0.75\columnwidth}{!}{%
\includegraphics{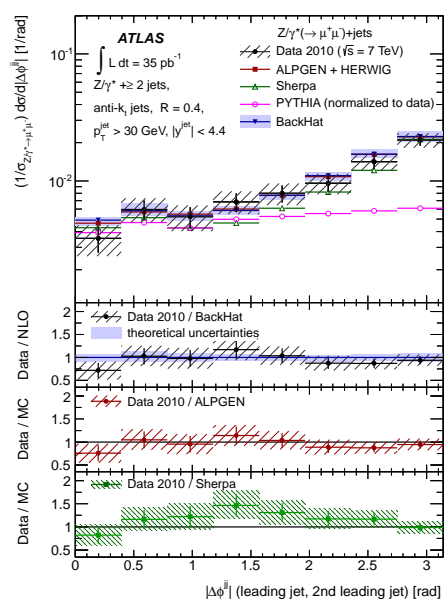} }
\end{center}
\caption{Dijet cross section in Z/$\gamma^*(\to \mu^+\mu^-)$+jets production normalized by Drell-Yan
cross section as a function of the azimuthal separation of the two leading jets.}
\label{fig:4}
\end{figure}

\begin{figure}
\begin{center}
\resizebox{0.75\columnwidth}{!}{%
\includegraphics{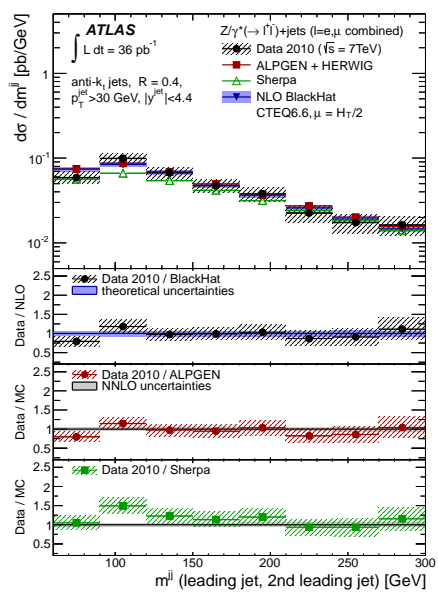} }
\end{center}
\caption{Dijet cross section in Z/$\gamma^*(\to l^+l^-)$+jets production as a function of the invariant
mass of the two leading jets.}
\label{fig:5}
\end{figure}

Fig. \ref{fig:1} presents the measured fiducial cross section as a function of the inclusive jet multiplicity 
in events with up to at least four jets in the final state for the electron channel.\\
The measured ratio of cross sections for N$_{jet}$ and N$_{jet}$-1 in the muon channel is shown in Fig. \ref{fig:2}.
This observable cancels part of the systematic uncertainty and constitutes an improved test of the Standard Model.
The data indicate that the
cross sections decrease by a factor of approximately five with the requirement of each additional jet in the final state.\\
The inclusive jet differential cross section  as a function of p$_T$ is presented in Fig. \ref{fig:3} 
  in events with at least one jet in the final state for the electron channel.
The cross sections are divided by the corresponding
inclusive Z cross section times branching ratio with the aim
of canceling systematic uncertainties related to lepton identification and the luminosity.
The cross sections decrease by more than two orders of
magnitude as p$_T$ increases in the explored range.
Fig. \ref{fig:4}  shows the measured differential cross section as a function of the rapidity separation of the jets for the muon channel.\\
%
%
The combination of electron and
muon results has also been performed by extrapolating the measurements of the two channels to
a common lepton kinematical
region : p$_T$ $>$ 20 GeV and $ |\eta|<$ 2.5
as defined at the  vertex of the Z boson.
The electron and muon results are combined using the BLUE (Best Linear Unbiased Estimate)~\cite{ref:BLUE}  method.
Fig. \ref{fig:5}  shows the dijet cross section a function of the invariant
mass of the two leading jets.\\ 
The measured cross sections are in general well
described by NLO pQCD predictions including non-perturbative corrections, as well as by predictions of LO matrix
elements of up to 2 $\to$ 5 parton scatters, supplemented by parton showers, as implemented in the ALPGEN and
Sherpa MC generators.
In the case of PYTHIA, the LO pQCD ($q\bar{q}\to Z/\gamma^*g$ and $q g\to Z/\gamma^*q$ processes) MC predictions  underestimate
the measured cross sections.

\end{document}